\documentclass[twocolumn,showpacs,preprintnumbers,prl]{revtex4}%
\usepackage{amsfonts}
\usepackage{amsmath}
\usepackage{amssymb}
\usepackage{graphicx}%
\begin{document}
\title{Charge-Hall effect driven by spin force: reciprocal of the spin-Hall effect}
\author{Ping Zhang and Qian Niu}
\affiliation{Department of Physics, The University of Texas at Austin, Austin, TX 78712}

\begin{abstract}
A new kind of charge-Hall effect is shown. Unlike in the usual Hall effect,
the driving force in the longitudinal direction is a spin force, which may
originate from the gradient of a Zeeman field or a spin-dependent chemical
potential. The transverse force is provided by a Berry curvature in a mixed
position-momentum space. We can establish an Onsager relation between this
effect and the spin-Hall effect provided the spin current in the latter is
modified by a torque dipole contribution. This remarkable relation leads to
new ways for experimental detection of spin accumulation predicted by the spin
Hall effect.

\end{abstract}
\maketitle

The generation of an electric current in the transverse direction of an
electric field is known as a Hall effect. The transverse force is usually
provided by the Lorentz force from a magnetic field, but can also arise from
spin-orbit interactions in magnetic materials\cite{Chien} or Magnus forces on
vortices in superconductors\cite{Geller}. Apart from such ordinary and
anomalous Hall effects, other variants, such as the thermal Hall
effect\cite{Nolas}, can manifest when the longitudinal electric field is
replaced by other type of electromotive forces like a temperature gradient.
Likewise, a longitudinal electric field can also cause non-electrical response
in the transverse direction, such as in the spin Hall effect where a
transverse spin current may be generated\cite{Dya,Hirsch,Zhang}.

Recently, extensive theoretical work\cite{Jun,Fang,Yao,Ye} has established the
importance of intrinsic origin of the anomalous Hall effect in ferromagnets.
Unlike the traditional mechanisms of skew scattering\cite{Smit}, Berry phase
in the Bloch bands due to spin-orbit coupling can lead to a Hall current
depending only on the equilibrium part of the carrier distribution function.
In a similar fashion, intrinsic spin-Hall effect has been proposed for
paramagnetic materials\cite{Muk1,Sinova}, which has generated much interest
recently\cite{Culcer,Rashba,Hu,Sch1,Burkov,Bern,Inoue,Shen,Xiong,Usaj} in
association with electrical manipulation of spins for spintronics
applications\cite{Wolf,Awsch,Das}.

In this Letter, we propose a novel charge Hall effect driven by a spin force
along the longitudinal direction. This spin force can be provided by the
gradient of a Zeeman field or a spin dependent chemical potential. In the
intrinsic regime considered here, the transverse force is provided by a Berry
curvature in the mixed position-momentum space, also stemming from spin-orbit
coupling in the band structure. As shown below, there is not a longitudinal
electric current accompanying the spin-force in this intrinsic regime, so the
spin force should not be interpreted as a variant electromotive force.
Therefore, this effect is essentially different from the conventional thermal
Hall effect, which is known to be just a usual Hall effect driven by a variant
electromotive force.

Another goal of this Letter is to construct a "reciprocal" version of the
intrinsic spin-Hall
effect\cite{Muk1,Sinova,Culcer,Rashba,Hu,Sch1,Burkov,Bern,Inoue,Shen,Xiong,Usaj}
which is at present being in active investigation. Since a longitudinal charge
force (electric field) may drive a spin-Hall current through the spin-orbit
interaction, one naturally expects that a spin force can also induce a
charge-Hall current. To our surprise, we discovered that an exact Onsager
relation between these intrinsic Hall effects cannot readily be established;
only when the spin current is modified by including a torque dipole term, is
the spin-Hall conductivity in response to an electric field found to
correspond to our charge-Hall conductivity in response to a spin force.
Interestingly, it is this modified spin current that is responsible for spin
accumulation at a sample boundary as shown recently based on the spin
continuity equation in a semiclassical theory\cite{Culcer}. Our finding on the
Onsager relation further points to the importance of this modified spin
current. Spin accumulation due to spin Hall effect can thus be tested through
the Onsager relation by a measurement of our charge Hall effect in response to
a spin force.

To construct the theory, we consider the conduction electrons in semiconductor
quantum wells. In this two-dimensional electron gas (2DEG) system, spin-orbit
interaction arises from the quantum well asymmetry in the growth ($z$)
direction and has a standard Rashba form\cite{Rashba2} $V_{so}=(\alpha
/\hslash)(\sigma\times\mathbf{k})\cdot\mathbf{\hat{z}}$, where $\sigma$ is the
Pauli matrix vector and $\mathbf{k}$ is the 2D wave vector in the $x$-$y$
plane. The Rashba coefficient, $\alpha$, can be tuned over a wide range by a
vertical electric field. Taking into account the kinetic energy and a Zeeman
field normal to the heterostructure, we can write the $\mathbf{k}%
\cdot\mathbf{p}$ Hamiltonian as
\begin{equation}
H=\gamma k^{2}+V_{so}-\mu\sigma_{z}, \tag{1}%
\end{equation}
where $\gamma=\hslash^{2}/2m^{\ast}$ with $m^{\ast}$ being the band effective
mass, and $\mu$ is the Zeeman field (in unit of energy), which is assumed to
be weak in this paper. The orbital effect of the magnetic field can be taken
into account in our semiclassical formalism, but will be neglected in this
work for simplicity. To introduce a spin force, we consider a non-uniform
Zeeman field as in the Stern-Gerlach experiment for separating spin-up and
spin-down electrons. We thus assume $\mu=\mu_{0}+\mu_{1}x$, where $\mu_{0}$
describes the average strength of the Zeeman field while $F_{s}=(\frac
{2}{\hslash})\partial\mu/\partial x=\frac{2}{\hslash}\mu_{1}$ gives the spin
force in the $x$ (longitudinal) direction. The Zeeman field can be created by
exchange interaction with the moments of doped magnetic ions, which can be
polarized by a weak magnetic field (with negligible Lorentz force) at low
temperatures. An inhomogeneous Zeeman field can be produced either by a
magnetic field gradient or a temperature gradient. Alternatively, one may
consider using the spin-dependent chemical potential gradient present near an
interface with a ferromagnetic material.

To access transport properties, we adopt the formalism of semiclassical
wavepacket dynamics\cite{Chang}, which is a powerful tool for studying the
influence of slowly varying perturbations such as the spin force term in
Eq.(1) on the dynamics of Bloch electrons. Consider a wave packet centered at
$\mathbf{r}_{c}$ and with a narrow distribution around the mean wave vector
$\mathbf{k}$. Then the semiclassical equations of motion for $\mathbf{r}_{c}$
and $\mathbf{k}$ are \cite{Chang}%
\begin{align}
\mathbf{\dot{r}}_{c}  &  =\frac{1}{\hslash}\frac{\partial\varepsilon_{n}%
}{\partial\mathbf{k}}+\frac{1}{\hslash}\frac{\partial\Delta\varepsilon_{n}%
}{\partial\mathbf{k}}-\mathbf{\Omega}_{n}^{\mathbf{kr}}\cdot\mathbf{\dot{r}%
}_{c}-\mathbf{\Omega}_{n}^{\mathbf{kk}}\cdot\mathbf{\dot{k},}\tag{2a}\\
\mathbf{\dot{k}}  &  =-\frac{1}{\hslash}\frac{\partial\varepsilon_{n}%
}{\partial\mathbf{r}_{c}}-\frac{1}{\hslash}\frac{\partial\Delta\varepsilon
_{n}}{\partial\mathbf{r}_{c}}+\mathbf{\Omega}_{n}^{\mathbf{rr}}\cdot
\mathbf{\dot{r}}_{c}+\mathbf{\Omega}_{n}^{\mathbf{rk}}\cdot\mathbf{\dot{k},}
\tag{2b}%
\end{align}
where $n$ is the band index, and $\varepsilon_{n}$ is the local band energy
obtained by diagonalizing the Hamiltonian (1) with the position operator
$\mathbf{r}$ identified with the wavepacket center $\mathbf{r}_{c}$. The
result is $\varepsilon_{1,2}(\mathbf{k,r}_{c})=\gamma k^{2}\mp\sqrt{\mu
_{c}^{2}+\alpha^{2}k^{2}}$ with $\mu_{c}=\mu_{0}+\mu_{1}x_{c}$. In the second
term in Eqs.(2), $\Delta\varepsilon_{n}$ is the energy correction in the
gradient of the Zeeman field. For the Hamiltonian (1) we get $\Delta
\varepsilon_{1}(\mathbf{k,r}_{c})=\Delta\varepsilon_{2}(\mathbf{k,r}%
_{c})=\alpha^{2}k_{y}\mu_{1}/2(\mu_{c}^{2}+\alpha^{2}k^{2})$. Finally, the
last two terms in Eqs.(2) denote Berry-curvature corrections to the orbital
motion of the electron. They are defined by%
\begin{equation}
\left(  \mathbf{\Omega}_{n}^{\mathbf{kr}}\right)  _{\alpha\beta}\equiv
\Omega_{n}^{k_{\alpha}x_{\beta}}=i\left[  \langle\frac{\partial u_{n}%
}{\partial k_{\alpha}}|\frac{\partial u_{n}}{\partial x_{\beta}}%
\rangle-\langle\frac{\partial u_{n}}{\partial x_{\beta}}|\frac{\partial u_{n}%
}{\partial k_{\alpha}}\rangle\right]  , \tag{3}%
\end{equation}
where $u_{n}$ is the eigenstate of the $n$th band. The other Berry curvature
tensors are defined similarly. One can see that the equations of motion
involve Berry curvatures between every pair of parameters in combined real and
momentum spaces. Recent studies\cite{Jun, Fang,Yao} of the anomalous Hall
effect in ferromagnetic materials have revealed the fundamental role of the
Berry curvature\ $\mathbf{\Omega}^{\mathbf{kk}}$ in the $k$ space. As we will
see below, it is the Berry curvature in the mixed position-momentum space,
$\mathbf{\Omega}^{\mathbf{kr}}$, which has not attracted attention in previous
work, that determines the charge-Hall effect driven by a spin force.

In most cases, calculation of the Berry curvatures is an involved work. For
real materials with complicated crystal and ferromagnetic structures,
first-principle calculations are needed. Fortunately, in the present two-band
case, the Berry curvatures can be obtained analytically. For the bottom band
with energy $\varepsilon_{1}$, the non-zero components of the Berry curvatures
are
\begin{equation}
\Omega^{k_{x},k_{y}}=-\frac{\alpha^{2}\mu_{c}}{2\Delta_{c}^{3}},\Omega
^{k_{x},x}=\frac{\alpha^{2}k_{y}\mu_{1}}{2\Delta_{c}^{3}},\Omega^{k_{y}%
,x}=-\frac{\alpha^{2}k_{x}\mu_{1}}{2\Delta_{c}^{3}}, \tag{4}%
\end{equation}
where $\Delta_{c}=\sqrt{\mu_{c}^{2}+\alpha^{2}k^{2}}$. The Berry curvatures
for the top band are different from the above by a sign. Considering these
non-zero Berry curvatures, and keeping terms up to first order in spin force,
we get $\dot{k}_{x}=-\partial\varepsilon_{n}{\Large /}\hslash\partial x_{c}$
and $\dot{k}_{y}=0$. The orbital motion in the $n$th band is given by
\begin{align}
\dot{x}_{c}  &  =(1-\Omega_{n}^{k_{x},x})v_{n}^{x}+\frac{\partial
\Delta\varepsilon_{n}}{\hslash\partial k_{x}},\tag{5a}\\
\dot{y}_{c}  &  =v_{n}^{y}+\frac{\partial\Delta\varepsilon_{n}}{\hslash
\partial k_{y}}-\Omega_{n}^{k_{y},x}v_{n}^{x}+\Omega_{n}^{k_{y},k_{x}}%
\frac{\partial{\Large \varepsilon}_{n}}{\hslash\partial x_{c}}, \tag{5b}%
\end{align}
where $v_{n}^{\alpha}=(1/\hslash)\partial\varepsilon_{n}{\Large /}\partial
k_{\alpha}$ is band group velocity. The longitudinal and transverse currents
are obtained by averaging $\dot{x}_{c}$ and $\dot{y}_{c}$ with the
distribution function, respectively. The full distribution function consists
of an equilibrium part and a nonequilibrium part (maintained by a competition
between $k_{x}$-drift and scattering relaxation). As for the anomalous Hall
effect and spin Hall effect, there is an intrinsic contribution from the
equilibrium part of the distribution. The extrinsic contribution from the
non-equilibrium part will be small if scatterings are strong enough to
maintain the distribution near to equilibrium and weak enough (compared to the
Rashba splitting in energy scale) so that interband couplings are small. In
this work we will focus attention to the intrinsic regime in order to make
contact with similar work for the spin Hall effect. Using the equilibrium
distribution, we find that the longitudinal current vanishes. However, the
intrinsic Hall current in the $y$ direction is not zero, and is given by,
using Eq.(5b),
\begin{align}
J_{y}^{c}  &  =e\sum_{n=1}^{2}\int\frac{d^{2}k}{(2\pi)^{2}}f_{n}\Omega
_{n}^{k_{y},x}v_{n}^{x}\tag{6}\\
&  =-F_{s}\int\frac{d^{2}k}{(2\pi)^{2}}\frac{e\hslash\alpha^{2}k_{x}}%
{4\Delta^{3}}(f_{1}v_{1}^{x}-f_{2}v_{2}^{x})\nonumber\\
&  \equiv\sigma_{yx}^{cs}F_{s},\nonumber
\end{align}
where $f_{n}$ is the equilibrium Fermi-Dirac distribution function, and
$\Delta=(\mu_{0}^{2}+\alpha^{2}k^{2})^{1/2}$. We note that the contributions
from the second and last terms in Eq.(5b) cancel each other under the $k$
integral, thus only the Berry curvature $\Omega^{k_{y},x}$ contributes to the
charge-Hall current.%

%TCIMACRO{\TeXButton{TeX field}{\begin{figure}[tbp]
%\begin{center}
%\includegraphics[width=1.0\linewidth]{fig1.eps}
%\end{center}
%\caption{Charge-Hall conductivity (in unit of $e$) in GaAs quantum well system
%as a function of (a) Rashba coefficient $\alpha$ and the electron density
%$N_{e}$ (contour plot) for $T=0$ and $\mu_{0}=0.01$meV, (b) Rashba coefficient
%for different values of $\mu_{0}$, and (c) temperature.}
%\end{figure}}}%
%BeginExpansion
\begin{figure}[tbp]
\begin{center}
\includegraphics[width=1.0\linewidth]{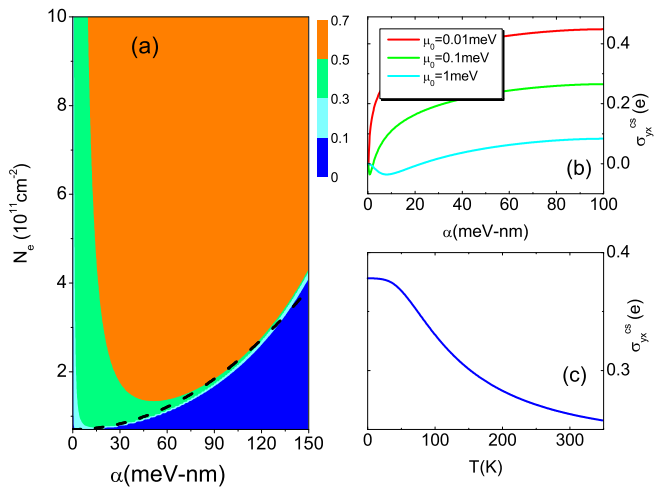}
\end{center}
\caption{Charge-Hall conductivity (in unit of $e$) in GaAs quantum well system
as a function of (a) Rashba coefficient $\alpha$ and the electron density
$N_{e}$ (contour plot) for $T=0$ and $\mu_{0}=0.01$meV, (b) Rashba coefficient
for different values of $\mu_{0}$, and (c) temperature.}
\end{figure}%
%EndExpansion

Equation (6) is a central result in this paper. Figure 1 summarizes the
behavior of the charge-Hall conductivity. Panel (a) is a (colored) contour
plot of this quantity in the parameter plane of electron density and Rashba
coefficient. The (blue) region below the dashed line corresponds to the case
of single band occupation, where the charge-Hall conductivity is small. At
high electron densities, $\sigma_{yx}^{cs}$ remains large within a wide range
of $\alpha$. Panel (b) shows the dependence on the Rashba coefficient for a
set of Zeeman energies. In these panels, the temperature is set to zero. Panel
(c) shows that the charge-Hall effect is robust against thermal broadening, a
property stemming from the intrinsic nature of $\sigma_{yx}^{cs}$. For
estimate, we consider a GaAs quantum well sample with electron density
$N_{e}\sim5\times10^{11}$cm$^{-2}$ and a typical Rashba coefficient of
$\alpha\sim20$meV-nm. A spin force corresponding to $\mu_{1}\sim1$meV/$\mu$m
(one millivolt of Zeeman energy change over a micron) can yield a charge-Hall
current on the order of $1$mA/cm, implying that the predicted effect is observable.

Now we turn our attention to a basic and enlighten question: is there an
Onsager relation between our spin force-driven charge-Hall effect and the
electric field-driven spin-Hall effect? In the presence of an electric field
along the $y$ direction, the Hamiltonian is now%
\begin{equation}
H=\gamma k^{2}+V_{so}-\mu_{0}\sigma_{z}+eEy, \tag{7}%
\end{equation}
where we have removed the spin force but kept the uniform part of the Zeeman
field. Based on the spin current operator, defined as the symmetric product of
the velocity and spin operators, $\frac{1}{2}(s_{z}\dot{x}+\dot{x}s_{z})$, one
can obtain a spin current in the transverse ($x$) direction either from the
Kubo formula or the semiclassical wavepacket formalism. In the intrinsic
regime, this spin current for the Rashaba model is given by
\begin{equation}
J_{x}^{s}=-E\int\frac{d^{2}k}{(2\pi)^{2}}\frac{e\hslash^{2}\alpha^{2}k_{x}%
^{2}}{4m^{\ast}\Delta^{3}}(f_{1}-f_{2}). \tag{8}%
\end{equation}
At zero temperature and $\mu_{0}=0$, this yields a universal spin Hall
conductivity \cite{Sinova,Hu,Sch1,Burkov,Inoue,Shen,Xiong}, $\sigma_{xy}%
^{sc}=-\sigma_{yx}^{sc}=-e/8\pi$, which does not depend on the Rashba
coefficient. Unfortunately, this universality is absent in the spin
force-driven charge-Hall conductivity $\sigma_{yx}^{cs}$ [see Eq.(6) and
Fig.1]. In fact, the expression for $\sigma_{yx}^{cs}$ in Eq.(6) can be
reduced to the expression for $\sigma_{xy}^{sc}$ in Eq.(8) \textit{only} when
the band group velocity $v_{1,2}^{x}=\hslash k_{x}/m^{\ast}\mp\alpha
^{2}/\hslash\Delta^{2}$ is replaced by its first term $\hslash k_{x}/m^{\ast}%
$; the neglected term will cause a finite difference for non-zero Rashba
coefficient. Thus contrary to our intuitive expectation, a full Onsager
relation between these two kinds of Hall conductivities fails to be established.

To understand the origin of this violation of the Onsager relation and to
search for a remedy, we notice that the spin-current operator is not really an
observable in conjugation with the spin force. To illustrate this point, let
us put the charge force (the electric field) and the spin force in an equal
footing by including both of them in the perturbative Hamiltonian%
\begin{equation}
\delta H=-F_{1}d_{1}-F_{2}d_{2}, \tag{9}%
\end{equation}
where $F_{1}=E$ and $F_{2}=F_{s}$ are the generalized forces applied on the
charge and spin degrees of freedom, whereas $d_{1}=-ey$ and $d_{2}=s_{z}x$ are
the corresponding displacement operators. The response currents are obtained
as expectation values of the generalized velocity operators in the perturbed
state
\begin{align}
\dot{d}_{1}  &  =-e\dot{y},\tag{10a}\\
\dot{d}_{2}  &  =s_{z}\dot{x}+\dot{s}_{z}x\mathbf{,} \tag{10b}%
\end{align}
where symmetrization between operators in Eq.(10b) is implied. It is clear now
that the current conjugate to the spin force ($F_{2}$) not only involves the
spin current given from $s_{z}\dot{x}$\cite{Muk1,Sinova}, but also a torque
dipole term $\dot{s}_{z}x$. We therefore introduce a modified spin current
defined by
\begin{equation}
\tilde{J}_{x}^{s}=\sum_{n}\int\frac{d^{2}k}{(2\pi)^{2}}f_{n}\langle s_{z}%
\dot{x}+\dot{s}_{z}x\rangle_{n}, \tag{11}%
\end{equation}
where the bracket indicates quantum mechanical average over the wavepacket
made out of the $n$th band. Due to the spin-orbit interaction, besides driving
the wavepacket in the $\mathbf{k}$-space, the electric field will also give
rise to a nonadiabatic correction to the spin wavefunction. This correction
has been taken into account when constructing our wavepacket for each band.

In terms of the wave packet center coordinate $x_{c}$ and velocity $\dot
{x}_{c}$, we can rewrite the above expression as \cite{Culcer}%
\begin{equation}
\tilde{J}_{x}^{s}=\sum_{n}\int\frac{d^{2}k}{(2\pi)^{2}}f_{n}\left[  \dot
{x}_{c}\langle s_{z}\rangle_{n}+{\frac{d}{dt}(}p_{x}^{s_{z}}{)}_{n}%
+x_{c}{\frac{d}{dt}}\langle s_{z}\rangle_{n}\right]  , \tag{12}%
\end{equation}
where ${(}p_{x}^{s_{z}}{)}_{n}=\langle(x-x_{c})s_{z}\rangle_{n}$ is the
$x$-component of the spin dipole in the wave packet. To first order in the
electric field, the time derivatives in the second and third terms can be
replaced by $-eE(\partial/\hslash\partial k_{y})$, and the quantities under
the time derivatives can be evaluated at zero field as $\langle s_{z}%
\rangle_{1,2}=\pm\mu_{0}\hslash/2\Delta$, and ${(}p_{x}^{s_{z}}{)}%
_{1,2}=-\hslash\alpha^{2}k_{y}/4\Delta^{2}$. It is then clear from symmetry
that the last term in the above equation vanishes after the $k$ integral. The
wave packet velocity has a first order field correction due to Berry phase,
$\dot{x}_{c}=v_{n}^{x}+(e/\hslash)E\Omega_{n}^{k_{x}k_{y}}$, where $v_{n}%
^{x}=\partial\varepsilon_{n}/\hslash\partial k_{x}$ is the band group
velocity. The Berry curvature tensor $\mathbf{\Omega}_{n}^{\mathbf{kk}}$ has
been given in Eq.(4). Similarly, the wave packet spin also has a first order
field correction, given by $E\langle s_{z}\rangle_{n}^{\prime}$ with $\langle
s_{z}\rangle_{1,2}^{\prime}=\mp e\alpha^{2}k_{x}/4\Delta^{3}$. After these
considerations, the modified spin-Hall current can be written as%
\begin{align}
\tilde{J}_{x}^{s}  &  =E\sum_{n}\int\frac{d^{2}k}{(2\pi)^{2}}f_{n}[v_{n}%
^{x}\langle s_{z}\rangle_{n}^{\prime}+\frac{e}{\hslash}\Omega_{n}^{k_{x}k_{y}%
}\langle s_{z}\rangle_{n}\tag{13}\\
&  -\frac{e}{\hslash}\frac{\partial}{\partial k_{y}}{(}p_{x}^{s_{z}}{)}%
_{n}].\nonumber
\end{align}
Remarkably, we find that the contributions from the Berry curvature term and
the spin dipole term completely cancel each other. Thus only the first term in
Eq.(13) contributes to the modified spin current, yielding a spin Hall
conductivity
\begin{equation}
\sigma_{xy}^{sc}=-\int\frac{d^{2}k}{(2\pi)^{2}}\frac{e\hslash\alpha^{2}k_{x}%
}{4\Delta^{3}}(f_{1}v_{1}^{x}-f_{2}v_{2}^{x}). \tag{14}%
\end{equation}

A comparison between Eq.(14) and Eq.(6) gives the Onsager relation
\begin{equation}
\sigma_{yx}^{cs}(\mu_{0})=\sigma_{xy}^{sc}(\mu_{0}). \tag{15}%
\end{equation}
It should be pointed out that these two kinds of Hall conductivities have
time-reversal symmetry $T$, which is absent in the conventional Ohm's law.
Since the spin force and the charge-Hall current are odd under $T$, thus they
can be related via $T$-invariant $\sigma_{yx}^{cs}$. On the other hand, the
spin current and the electric field are even under $T$; the resultant
spin-Hall conductivity $\sigma_{xy}^{sc}$ is also $T$-invariant. Due to this
time-reversal symmetry, the general Onsager relation $\sigma_{\alpha\beta}%
(\mu_{0})=\sigma_{\beta\alpha}(-\mu_{0})$ reduces to Eq.(15) in the present
context. We have also shown explicitly\cite{Ping} that Eq.(15) remains to be
true for the 4-band Luttinger model, and we believe that this is a generally
valid relationship.

This remarkable Onsager relation is another central result in this paper. As a
consequence, the spin Hall conductivity (for the modified spin current) should
behave the same as we depicted in Fig. 1 for our charge Hall conductivity in
response to a spin force. All previous discussions on spin Hall effect
measurement are based on spin accumulation on a sample boundary due to a bulk
spin current; Ref.\cite{Culcer} showed that it is the modified spin current
that is relevant to such a measurement. The Onsager relation derived here
provides another method for testing this modified spin Hall effect.

In conclusion, we have shown that a spin force can drive a novel charge-Hall
current through the spin-orbit coupling. The predicted charge Hall effect has
a remarkable Onsager relation with a spin-Hall effect driven by an electric
field. This Onsager relation is only attainable when the spin current is
corrected by a torque dipole density.

We thank D. Culcer, J. Sinova, and A.H. MacDonald for useful discussions. PZ
was supported by NSF Grand No. DMR-0071893 and Welch Foundation in Texas. QN
was supported by DE-FG03-02ER45958.

\end{document}